\documentclass{aa}
\usepackage{epsfig}
\begin{document}
%
\title{Nova-Like Cataclysmic Variable TT Ari:}
 \subtitle{QPO Behaviour Coming Back From Positive Superhumps
\fnmsep\thanks{Tables 1,2,4 and Figures 7,10 are only available in electronic form at http://www.edpsciences.org}
}
   \author{Y. Kim\inst{1,2} \and I.L. Andronov\inst{3,4}
\and S.M.Cha\inst{1} \and L.L.Chinarova\inst{4} \and J.N. Yoon\inst{1}
         }
   \institute{
University Observatory, Chungbuk National University, 361-763,
Cheongju, Korea
         \and
Institute for Basic Science Research, Chungbuk National
University, 361-63, Korea
         \and
Odessa National Maritime University,
             Mechnikov str., 34, 65029, Odessa, Ukraine, \email{tt\_ari@ukr.net}
         \and
Astronomical Observatory, Odessa National University,
             T.G.Shevchenko Park, 65014, Odessa, Ukraine
             }
\date{Received April 18, 2008, Revised September 21, 2008 / Accepted September 29, 2008}
\abstract{}{We study the variability of the nova-like cataclysmic variable TT Ari, on time-scales of between minutes and months.}
 {The observations in the filter R were obtained at the 40-cm telescope of the Chungbuk National University (Korea), 51 observational runs cover 226 hours. The table of individual observations is available electronically. For the analysis, we have used several methods: periodogram, wavelet and scalegram analysis.} {TT Ari was in the ``negative superhump" state after its return from the ``positive superhump" state, which lasted 8 years. The ephemeris for 12 best pronounced minima is $T_{min}=BJD 2453747.0700(47)+0.132322(53)E.$
The phases of minima may reach 0.2, indicating non-eclipse nature of these minima. The quasi-periodic oscillations (QPO) are present with a mean ``period" of 21.6 min and mean semi-amplitude of 36 mmag.
This value is consistent with the range 15-25 minutes reported for previous ``negative superhump" states and does not support the hypothesis of secular decrease of the QPO period.
 Either the period, or the semi-amplitude show significant night-to-night variations. According to the position at the two-parameter diagrams (i.e. diagrams of pairs of parameters: time, mean brightness of the system, brightness of the source of QPO, amplitude and time scale of the QPOs), the interval of observations was splitted into 5 parts, showing different characteristics: 1) the ``pre-outburst" stage, 2) the ``rise to outburst", 3) ``top of the outbursts", 4) ``post-outburst QPO" state, 5) ``slow brightening". The brightness of the QPO source  was significantly larger during the 10-day outburst, then for the preceding interval. However, after the outburst, the large brightness of the QPO source still existed for around 30 days, producing the stage ``4". The corresponding diagram $m_{QPO}(\bar{m})$ has a splitting into two groups at the brightness range 10\fm6-10\fm8, which correspond to larger and smaller amplitudes of the QPO.
For the group ``5" only,
statistically significant correlations were found, for which, with increasing mean brightness, also increase the period, amplitude and brightness of the source of QPOs.
The mean brightness at the ``negative superhump state" varies within 10\fm3-11\fm2, so the system is brighter than at the ``positive superhump" (11\fm3), therefore the ``negative superhump" phenomenon may be interpreted by a larger accretion rate.
The system is an excellent laboratory to study processes resulting in variations at time-scales from seconds to decades and needs further monitoring at various states of activity.
}{}
\authorrunning{Kim, Andronov, et al.
}
\titlerunning{TT Ari: QPO Behaviour Coming Back From Positive Superhumps}
\maketitle


\section{Introduction}

TT Ari is one of the brightest and thus extensively studied nova-like cataclysmic variable (see e.g. monographs: Warner (1995), Hellier (2001)). A wide variety of processes takes place in this object, making it variable at time scales from 9.6 seconds (Kozhevnikov, 1986)
to years (Hudec et al. 1984, Bianchini 1990, Kraicheva et al. 1999). Due to a low inclination, there are no prominent eclipses at the light curve, but there was a "negative superhump"-type modulation and $\sim20-$min quasi-periodic oscillations (QPO). Mardirossian et al. (1980) reported on $\sim40^{\rm s}$ "oscillatory events", $\ge6$-min QPOs, flickering and a nearly sinusoidal shape of $3-$ hour modulations.

Krautter et al. (1981) classified TT Ari as "a dwarf nova at a permanent outburst". Hoard (2007) included TT Ari in his "Big list" of the SW Sex stars (see e.g. Hellier 2000 for a review).
Semeniuk et al. (1987) suggested 4 periods in TT Ari, namely, 3.8 day beat period of the 3.2 - hour photometric and 3.3 hour spectroscopic periods, and quasi-periodic oscillations (QPO) with an apparent decrease of the QPO period from 27 (in 1961) to 17 minutes (in 1985). Kozhevnikov (1986) found variability at the short timescale 9.6 sec.

 To study these types of variability, several international observational campaigns have been organized, the results of which have been summarized by Wenzel et al. (1986), Tremko et al. (1996), Skillman et al. (1998) and Andronov et al. (1999). Hollander and van Paradijs (1992) suggested a dependence of the characteristics of QPOs on accretion rate. Kraicheva et al. (1999) suggested variations of the accretion rate as a cause of switches between the positive and negative superhumps. 


\begin{figure*}
\psfig{file=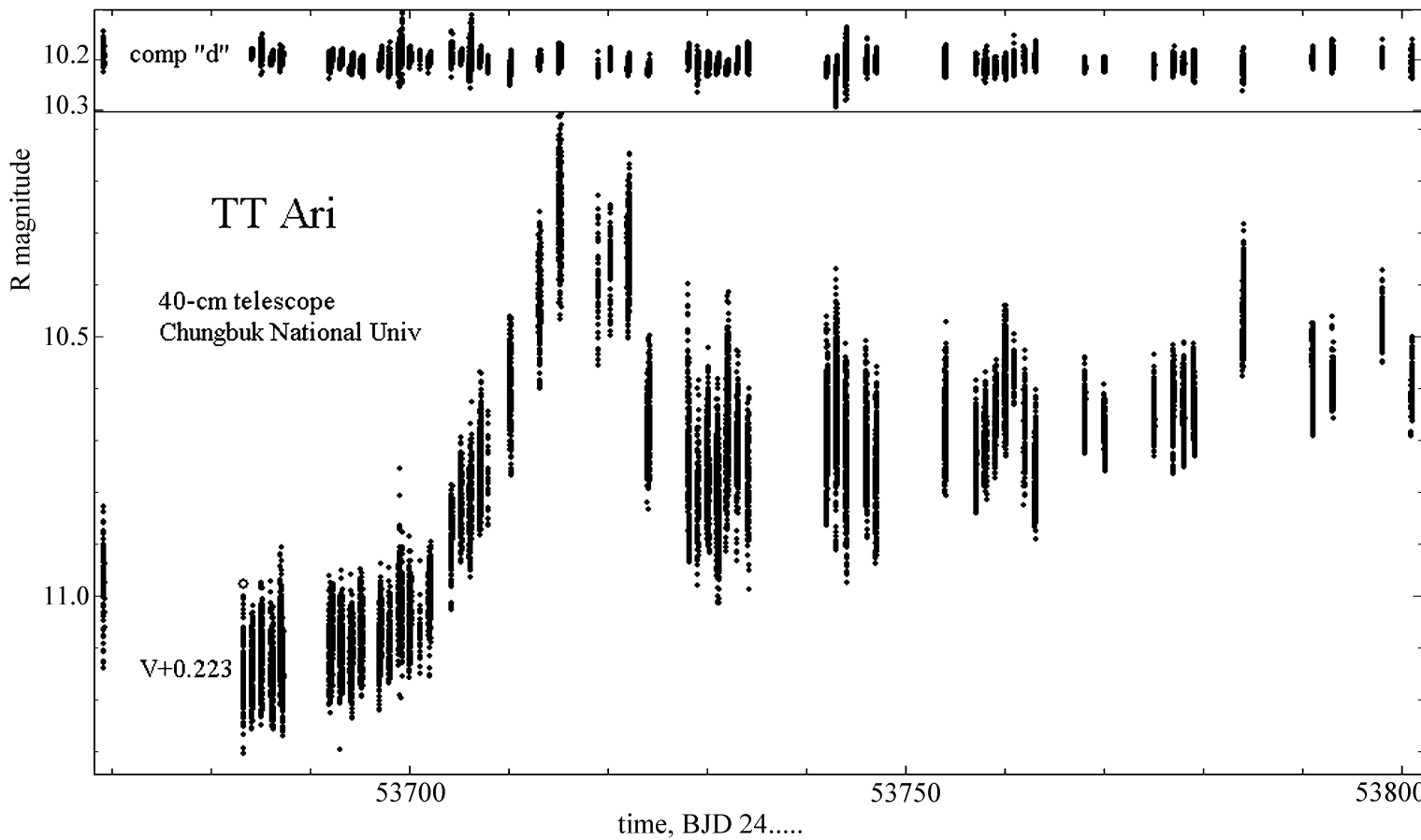} \label{f1}
\caption{The complete light curve
of TT Ari for 51 nights of observations (bottom) and of the
comparison star ``d". The V magnitudes in one night (53683) of
observations were shifted using the mean value of the instrumental
color index V-R=0\fm223.}
\end{figure*}

 One of the most exciting events in TT Ari was a switching in 1997 from the state of "negative" superhumps, which lasted nearly 3 decades of optical studies, to a state of "positive" superhumps (Skillman et al. 1998), in an excellent agreement with the $"P_{orb}-P_{sh}"$ ("orbital period - superhump period") statistical relation (Andronov et al. 1999). The system showed highly asymmetric (M-m=0.320(5)) permanent superhumps with a same amplitude of $0\fm19$ both in V and R and a period of $0\fd14841(1)$ in 2004 (Andronov et al. 2005). However, in 2005, the system has changed its state back to negative superhumps with possibly some additional evidence for instability (Andronov et al. 2005). It caused to start a regular photometric monitoring at the Chungbuk National University. Preliminary results were briefly published by Yoon et al. (2006). Results of this observational campaign are presented in this paper.

\begin{table*}
\label{t1}
\caption{(Available electronically only). Journal of observations of TT Ari:
The "legend Julian date LJD" (integer part of the Julian date for the beginning of the night); number of observations $n$; time of the begin $t_b$ and end $t_e$ of observations (for the many nights, the observations ended on the next integer JD); duration of the run $t_e-t_b$; magnitude range for individual data points $m_{max}$, $m_{min}$;
nightly mean $\bar{m}$ and it's accuracy estimate $\sigma(\bar{m})$;
r.m.s. deviation of the single observation from the mean $\sigma(m);$
exp - time resolution in seconds. All moments of time are expressed in (BJD-2400000).
}
\begin{tabular}{crccccccr}
\hline
LJD&~~$n$&$t_b-t_e$&duration&range& $\bar{m}$&$\sigma(\bar{m})$&$\sigma(m)$&exp\\
\hline
53668&   89& 53669.12931-69.20075& 0.07144& 10.847-11.167& 10.9975&  0.0073&  0.0687 &  64\\[-0mm]
53682&  127& 53683.16957-83.29230& 0.12273& 10.995-11.338& 11.1820&  0.0057&  0.0641 &  77\\[-0mm]
53683&  243& 53684.03965-84.24841& 0.20876& 11.045-11.289& 11.1667&  0.0033&  0.0508 &  62\\[-0mm]
53684&  213& 53684.97582-85.14497& 0.16915& 11.000-11.282& 11.1341&  0.0042&  0.0613 &  53\\[-0mm]
53685&  326& 53685.98716-86.29891& 0.31175& 10.997-11.288& 11.1767&  0.0028&  0.0509 &  67\\[-0mm]
53686&  445& 53686.91790-87.27436& 0.35646& 10.929-11.303& 11.1435&  0.0031&  0.0646 &  58\\[-0mm]
53691&  545& 53691.89585-92.26499& 0.36914& 10.986-11.256& 11.1194&  0.0021&  0.0491 &  48\\[-0mm]
53692&  371& 53692.94350-93.25535& 0.31185& 10.975-11.239& 11.1228&  0.0026&  0.0494 &  53\\[-0mm]
53693&  302& 53693.97067-94.27612& 0.30545& 10.982-11.268& 11.1418&  0.0029&  0.0497 &  78\\[-0mm]
53694&  331& 53694.98199-95.26401& 0.28202& 10.972-11.224& 11.1039&  0.0028&  0.0506 &  68\\[-0mm]
53696&  326& 53696.92302-97.16962& 0.24660& 10.962-11.252& 11.1090&  0.0028&  0.0506 &  49\\[-0mm]
53698&  376& 53698.88323-99.29020& 0.40697& 10.773-11.228& 11.0404&  0.0028&  0.0548 &  54\\[-0mm]
53700&   38& 53701.05339-01.09047& 0.03708& 10.956-11.180& 11.0932&  0.0075&  0.0460 &  73\\[-0mm]
53703&  159& 53704.14776-04.27062& 0.12286& 10.804-11.052& 10.9180&  0.0044&  0.0561 &  57\\[-0mm]
53704&  216& 53705.07537-05.24029& 0.16492& 10.712-10.959& 10.8413&  0.0035&  0.0509 &  54\\[-0mm]
53706&  391& 53707.03617-07.24017& 0.20400& 10.582-10.905& 10.7678&  0.0031&  0.0605 &  34\\[-0mm]
53707&   31& 53707.89166-07.92107& 0.02941& 10.660-10.885& 10.7859&  0.0106&  0.0593 &  69\\[-0mm]
53712&  301& 53713.00121-13.16906& 0.16785& 10.265-10.615& 10.4350&  0.0040&  0.0692 &  28\\[-0mm]
53714&  408& 53714.96644-15.24025& 0.27381& 10.071-10.478& 10.2661&  0.0036&  0.0735 &  54\\[-0mm]
53718&   53& 53718.93715-18.96768& 0.03053& 10.233-10.569& 10.4005&  0.0113&  0.0820 &  48\\[-0mm]
53719&   60& 53720.16045-20.20221& 0.04176& 10.251-10.509& 10.3689&  0.0082&  0.0637 &  59\\[-0mm]
53721&  463& 53721.90647-22.18562& 0.27915& 10.150-10.514& 10.3352&  0.0031&  0.0665 &  43\\[-0mm]
53723&  419& 53723.89227-24.20482& 0.31255& 10.510-10.854& 10.6845&  0.0031&  0.0630 &  49\\[-0mm]
53727&  325& 53728.01029-28.20304& 0.19275& 10.407-10.959& 10.7631&  0.0059&  0.1071 &  48\\[-0mm]
53728&  138& 53728.90736-29.15139& 0.24403& 10.597-11.003& 10.8262&  0.0065&  0.0759 & 119\\[-0mm]
53729&  418& 53729.88847-30.15052& 0.26205& 10.534-10.939& 10.7779&  0.0038&  0.0779 &  48\\[-0mm]
53730&  467& 53730.88198-31.16156& 0.27958& 10.596-11.038& 10.8290&  0.0039&  0.0848 &  48\\[-0mm]
53731&  457& 53731.87834-32.14991& 0.27157& 10.422-10.938& 10.6882&  0.0045&  0.0957 &  48\\[-0mm]
53733&  199& 53734.00208-34.14173& 0.13965& 10.615-11.013& 10.7913&  0.0053&  0.0745 &  59\\[-0mm]
53741&  426& 53741.93878-42.34369& 0.40491& 10.400-10.887& 10.6814&  0.0043&  0.0886 &  40\\[-0mm]
53742&  490& 53742.88202-43.12333& 0.24131& 10.378-10.933& 10.6474&  0.0042&  0.0938 &  41\\[-0mm]
53743&  302& 53743.88083-44.04232& 0.16149& 10.525-11.000& 10.7455&  0.0055&  0.0949 &  41\\[-0mm]
53753&  355& 53753.88591-54.07058& 0.18467& 10.482-10.827& 10.6851&  0.0035&  0.0665 &  41\\[-0mm]
53756&  201& 53756.97976-57.10862& 0.12886& 10.597-10.861& 10.7513&  0.0042&  0.0590 &  51\\[-0mm]
53757&  428& 53757.89002-58.07701& 0.18699& 10.582-10.836& 10.7254&  0.0022&  0.0463 &  35\\[-0mm]
53758&  403& 53758.89195-59.07778& 0.18583& 10.557-10.792& 10.6759&  0.0023&  0.0452 &  26\\[-0mm]
53759&  403& 53759.90846-60.08968& 0.18122& 10.450-10.748& 10.5982&  0.0033&  0.0659 &  26\\[-0mm]
53760&   87& 53760.89155-60.92635& 0.03480& 10.507-10.646& 10.5905&  0.0029&  0.0269 &  30\\[-0mm]
53761&  199& 53761.89168-61.98891& 0.09723& 10.540-10.846& 10.6917&  0.0041&  0.0572 &  41\\[-0mm]
53762&  305& 53762.92413-63.08970& 0.16557& 10.618-10.914& 10.7826&  0.0033&  0.0573 &  45\\[-0mm]
53767&  217& 53767.94171-68.05934& 0.11763& 10.552-10.744& 10.6707&  0.0027&  0.0401 &  46\\[-0mm]
53769&  228& 53769.90351-70.05308& 0.14957& 10.606-10.778& 10.7000&  0.0025&  0.0381 &  46\\[-0mm]
53774&  177& 53774.92833-75.04274& 0.11441& 10.548-10.748& 10.6563&  0.0030&  0.0403 &  29\\[-0mm]
53776&  144& 53776.90179-76.97372& 0.07193& 10.529-10.783& 10.6487&  0.0049&  0.0587 &  35\\[-0mm]
53777&  289& 53777.90157-78.05429& 0.15272& 10.523-10.770& 10.6635&  0.0029&  0.0498 &  35\\[-0mm]
53778&  357& 53778.91810-79.05101& 0.13291& 10.526-10.748& 10.6468&  0.0024&  0.0462 &  18\\[-0mm]
53783&  298& 53783.92239-84.03073& 0.10834& 10.289-10.590& 10.4454&  0.0036&  0.0619 &  29\\[-0mm]
53790&  218& 53790.91312-91.02148& 0.10836& 10.484-10.709& 10.5921&  0.0041&  0.0608 &  40\\[-0mm]
53792&  173& 53792.91504-93.02008& 0.10504& 10.471-10.673& 10.6038&  0.0025&  0.0331 &  45\\[-0mm]
53797&   91& 53797.95336-98.00170& 0.04834& 10.381-10.563& 10.4842&  0.0045&  0.0432 &  45\\[-0mm]
53800&  171& 53800.91434-00.98895& 0.07461& 10.513-10.709& 10.6096&  0.0032&  0.0424 &  20\\
\hline
\end{tabular}

\end{table*}

\begin{table}
\label{t2}
\caption{Table of R observations (HJD-2400000, magnitude) obtained at the 40cm telescope of the CBNU observatory, Korea. The table (14199 lines) is published electronically only.}
\end{table}




\section{Observations and Comparison Stars}

\subsection{Observations}

The time-series observations in R band were made using a 512$\times$512 FLI CM-1 CCD
attached to the 40 cm telescope at Chungbuk National University (MAEDE LX200). The
field of view of a CCD image is about 12.97$\times$12.97  arcmin$^2$, given a CCD
plate scale of  1."52 per pixel at the f/6.3 Schmidt-Cassegrain focus of the
telescope.

Instrumental signatures of each CCD frame were removed and calibrated using the
bias, dark, and flat field frames, with the aid of the IRAF package CCDRED. We
obtained instrumental magnitudes of stars from the empirical point-spread
function (PSF) fitting method in the IRAF package DAOPHOT (Stetson 1987; Massey
\& Davis 1992).

The journal of observations is presented in Table 1. All Julian dates are expressed omitting "24" at the beginning, so the integer part contains 5 digits instead of 7.
For the legend of a separate night of observations, we have used the integer "legend Julian Date" LJD=int(BJD$_1-0.5),$ where BJD is bary-centric Julian date, BJD$_1$ is BJD for the first observation, and int$(x)$ is an integer part of $x.$ An additional substraction of 0.5 was needed because the integer part of BJD changed during the night many times. This method prevents duplication of LJD for different runs. E.g. the run starting on 53684.04 will have a designation LJD=53683, whereas the next run starting at 53684.97 (LJD=53684). So BJD--LJD is elapsed time in days since
the legend Julian date, and may exceed unity. For all observational runs,  BJD--LJD ranged from 0.88 to 1.34.
Totally 14199 observations were obtained during 226 hours in 51 nights (JD 2453669-2453800).

The original observations (BJD, R magnitude) are presented in Table 2 (electronically only).
For two nights, there was CCD V photometry (one night a series of V, another night sequences VRVR... with alternatively changing filters).

\begin{figure*}
\psfig{file=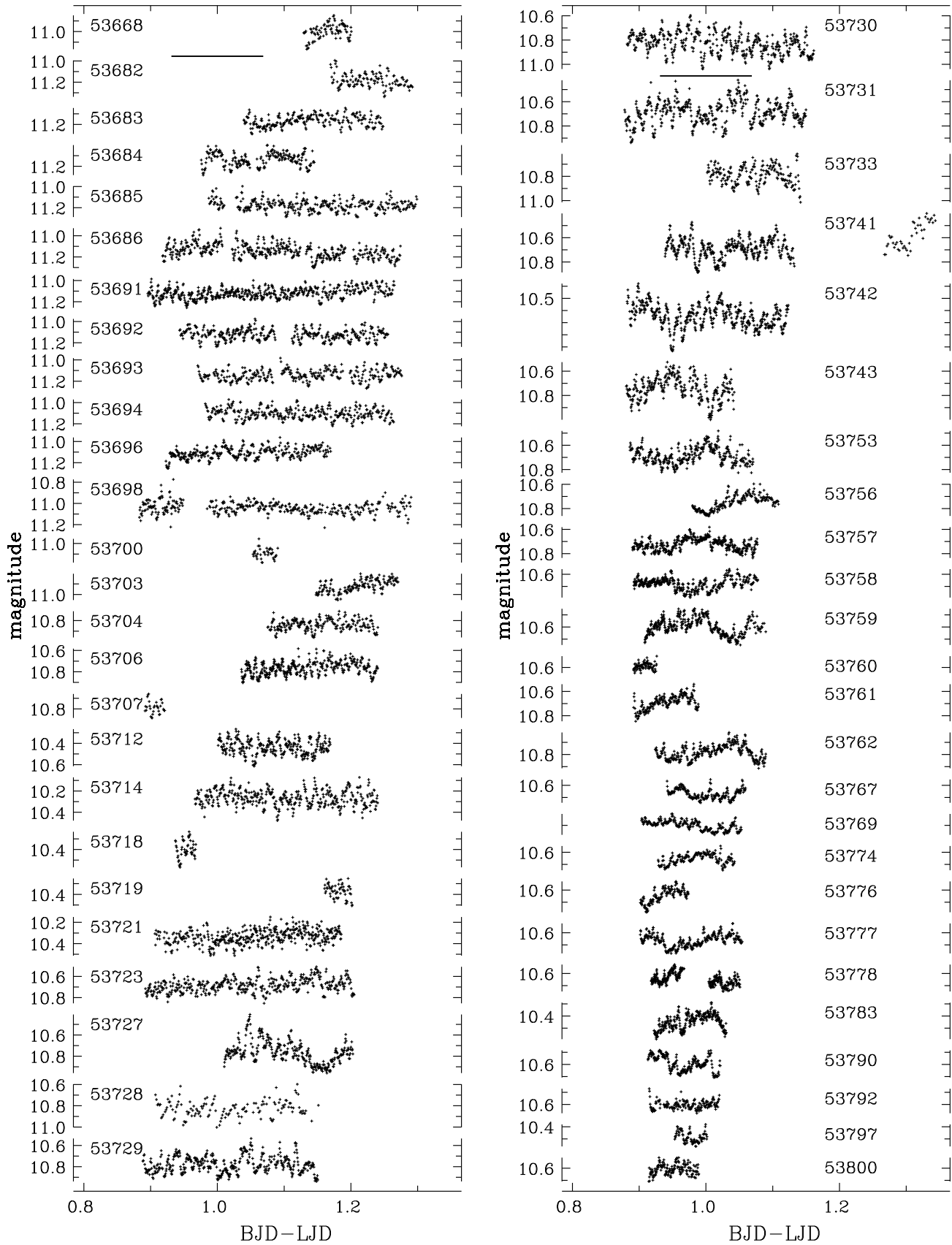} \label{f2}
\caption{The individual light curves of TT Ari. Abscissa is expressed as BJD-LJD, where the ``legend Julian date" LJD (-2400000) is shown near each curve. As the integer part of the Julian date JD often changed during our observations, we preferred to use the value corresponding to the beginning of the night. The length of horizontal lines between the first and second light curves is equal to the orbital period.}
\end{figure*}

\begin{table}
\label{t3}
\caption{Brightness of the comparison and check stars ("c" and "d", Goetz 1985) compiled from different publications.
}
\begin{tabular}{cccccr}
\hline
U&B&V&R&I&Ref.\\
\hline
\multicolumn{6}{c}{comparison star "c"= GSC 1207 1535}\\
12.04&11.72&11.01&10.41&9.97&Efimov et al. (1998)\\
11.91&11.68&10.99&&&Shafter et al. (1985)\\
-"-&-"-&-"-&&&Goetz (1985)\\
&11.74&11.05&10.66&10.3&Henden (2007)\\
&11.82&11.17&&&VSNET (2005)\\
&&11.1&&&AAVSO (2007)\\
\multicolumn{6}{c}{check star "d"= GSC 1207 1562}\\
13.38&12.17&11.02&&&Goetz (1985)\\
&12.20&11.05&10.51&9.97&Henden (2007)\\
&11.96&10.86&&&VSNET (2005)\\
&&11.1&&&AAVSO (2007)\\
\hline
\end{tabular}
\end{table}

\begin{table}
\label{t4}
\caption{(Available electronically only). Characteristics of the highest peaks at the ``wavelet periodogram" $S(P)$ for the individual nights: $\bar{t}$ - mean time of the run, $\bar{m}$ - sample mean magnitude (as in the Table 1), $m_{QPO}$ - stellar magnitude of QPOs, $\lg P,$  $\lg r,$ $S(P)$.
}
\begin{tabular}{cccccc}
\hline
$\bar{t}$&$\bar{m}$&$m_{QPO}$&$\lg P$ & $\lg r$ &$S(P)$\\
\hline
53669.16608& 10.9975& 13.8932& -2.06& -1.4237& 0.2054\\[-1mm]
53683.22808& 11.1820& 13.6733& -1.96& -1.2620& 0.4162\\[-1mm]
53684.13588& 11.1667& 14.1842& -1.94& -1.4724& 0.2807\\[-1mm]
53685.06281& 11.1341& 14.0890& -2.02& -1.4473& 0.2107\\[-1mm]
53686.14215& 11.1767& 14.2336& -2.01& -1.4881& 0.2177\\[-1mm]
53687.08587& 11.1435& 14.0220& -2.07& -1.4168& 0.2456\\[-1mm]
53692.06316& 11.1194& 14.2705& -2.03& -1.5258& 0.2309\\[-1mm]
53693.09523& 11.1228& 13.8781& -1.91& -1.3675& 0.3882\\[-1mm]
53694.12418& 11.1418& 13.9650& -1.99& -1.3947& 0.3655\\[-1mm]
53695.12250& 11.1039& 13.9996& -2.01& -1.4237& 0.2942\\[-1mm]
53697.03612& 11.1090& 14.3200& -1.86& -1.5498& 0.1978\\[-1mm]
53699.07330& 11.0404& 14.1175& -1.94& -1.4962& 0.1956\\[-1mm]
53701.07125& 11.0932& 13.9689& -2.04& -1.4157& 0.3457\\[-1mm]
53704.21028& 10.9180& 13.9387& -2.07& -1.4737& 0.3039\\[-1mm]
53705.15261& 10.8413& 13.9049& -1.97& -1.4908& 0.2235\\[-1mm]
53707.12664& 10.7678& 13.5723& -1.94& -1.3872& 0.2839\\[-1mm]
53707.90587& 10.7859& 12.9399& -1.98& -1.1273& 0.7236\\[-1mm]
53713.08494& 10.4350& 12.8533& -1.82& -1.2328& 0.3574\\[-1mm]
53715.10349& 10.2661& 13.0893& -1.92& -1.3947& 0.1543\\[-1mm]
53718.95241& 10.4005& 12.4503& -2.00& -1.0857& 0.5146\\[-1mm]
53720.18131& 10.3689& 13.3453& -1.98& -1.4559& 0.1677\\[-1mm]
53722.05972& 10.3352& 13.4575& -2.13& -1.5143& 0.1299\\[-1mm]
53724.04834& 10.6845& 13.7616& -1.75& -1.4962& 0.1488\\[-1mm]
53728.10643& 10.7631& 13.1026& -1.81& -1.2013& 0.2479\\[-1mm]
53729.01869& 10.8262& 13.1744& -1.80& -1.2048& 0.3683\\[-1mm]
53730.01934& 10.7779& 13.2913& -1.87& -1.2708& 0.2663\\[-1mm]
53731.02173& 10.8290& 13.1531& -1.87& -1.1952& 0.3237\\[-1mm]
53732.01320& 10.6882& 12.8525& -1.83& -1.1314& 0.3356\\[-1mm]
53734.07165& 10.7913& 13.5340& -1.90& -1.3625& 0.1931\\[-1mm]
53742.05811& 10.6814& 13.2361& -1.81& -1.2874& 0.2354\\[-1mm]
53743.00344& 10.6474& 12.9647& -1.81& -1.1925& 0.3037\\[-1mm]
53743.96000& 10.7455& 13.2835& -1.90& -1.2807& 0.2038\\[-1mm]
53753.97373& 10.6851& 13.7188& -2.02& -1.4789& 0.1812\\[-1mm]
53757.04199& 10.7513& 14.3629& -2.06& -1.7100& 0.1480\\[-1mm]
53757.98316& 10.7254& 14.3314& -2.08& -1.7077& 0.1520\\[-1mm]
53758.96880& 10.6759& 14.3272& -2.07& -1.7258& 0.1330\\[-1mm]
53759.98530& 10.5982& 13.9531& -1.96& -1.6073& 0.1150\\[-1mm]
53760.90784& 10.5905& 14.5697& -2.36& -1.8570& 0.1606\\[-1mm]
53761.94042& 10.6917& 13.9743& -1.83& -1.5784& 0.1907\\[-1mm]
53763.00687& 10.7826& 14.2102& -1.94& -1.6364& 0.1304\\[-1mm]
53768.00039& 10.6707& 14.3998& -1.93& -1.7570& 0.1229\\[-1mm]
53769.97754& 10.7000& 14.0952& -1.74& -1.6234& 0.3257\\[-1mm]
53774.98545& 10.6563& 14.1222& -1.93& -1.6517& 0.2238\\[-1mm]
53776.93634& 10.6487& 13.9733& -2.01& -1.5952& 0.1824\\[-1mm]
53777.97625& 10.6635& 14.2261& -2.09& -1.6904& 0.1596\\[-1mm]
53778.97798& 10.6468& 14.1078& -1.89& -1.6498& 0.1498\\[-1mm]
53783.97776& 10.4454& 13.5259& -1.84& -1.4976& 0.1797\\[-1mm]
53790.96543& 10.5921& 13.7395& -1.72& -1.5243& 0.1516\\[-1mm]
53792.97117& 10.6038& 14.0844& -1.87& -1.6576& 0.2388\\[-1mm]
53797.97754& 10.4842& 13.5245& -1.73& -1.4815& 0.2869\\[-1mm]
53800.95004& 10.6096& 14.0419& -1.72& -1.6383& 0.1495\\
\hline
\end{tabular}
\end{table}

\begin{table}
\label{t5}
\caption{Characteristics of the prominent minima of the ``3.2"-hour variability determined using the method of ``asymptotic parabola" (Marsakova and Andronov, 1996): time BJD (and its error) and value of the signal $m_min$ (and its error).}
\begin{tabular}{cccc}
\hline
t, BJD&$\sigma[t]$&$m_{min}$&$\sigma[m_{min}]$\\
\hline
53728.15648&  0.00180& 10.911&  0.014\\
53730.01153&  0.00065& 10.860&  0.009\\
53732.00013&  0.00122& 10.746&  0.009\\
53742.01621&  0.00078& 10.854&  0.014\\
53742.94917&  0.00234& 10.782&  0.019\\
53744.00722&  0.00220& 10.896&  0.023\\
53753.95263&  0.00001& 10.743&  0.007\\
53757.00094&  0.00140& 10.846&  0.005\\
53757.94716&  0.00072& 10.794&  0.008\\
53758.05255&  0.00162& 10.784&  0.010\\
53758.97022&  0.00248& 10.747&  0.008\\
53760.04089&  0.00030& 10.704&  0.005\\
\hline
\end{tabular}
\end{table}

\subsection{Comparison stars}
The photometric UBV standards in the field have been published by Goetz (1985) by linking to the star "c" (= GSC 1207 1535) of Shafter et al. (1985). The chart based on Hipparcos and Tycho catalogues is presented by the VSNET (2005) also in B and V. The UBVRI magnitudes were published only for one star "c", which were determined by Efimov et al. (1998) by linking to the standard star HD\,23949 (Neckel and Chini, 1980).
The check star "d" (= GSC 1207 1562), which is closer to TT Ari and suitable for measurements, is much redder (B-V=1\fm15) than "c" (B-V=0\fm69) (Goetz 1985), thus "c" is recommended to be used as the comparison star.

Recently, Henden (2007) published BVRI magnitudes for many stars in the field, including "c" (V=11\fm05, R=10\fm66) and "d".
A compilation of magnitude estimates is presented in Table 3.
The difference between the brightness estimates of the comparison star ``c" in the filter R reaches 0\fm25. This value is typical for discrepance between "standard" stellar magnitudes determined by different authors.
Finally, for the calibration of our observations, we adopted the values (V=11\fm01, R=10\fm41) published by Efimov et al. (1998). No color correction was made, so the instrumental magnitudes are defined as the adopted brightness of the comparison star plus instrumental difference "variable-comparison".
The error estimates for a single measurement of the variable are 0\fm008 (for the night at the top of the outburst) to 0\fm026 (for 6 nights in the faint state).

\section{Results of Analysis of Multi-Component Variability}

\subsection{Light Curves}

The multi-type character of variability was reported by previous researches. In this work, we make a time series analysis on an unprecedented number of observational runs, which were obtained in the same photometric system R (except 1 night in V). The complete light curve, obtained during the present observational campaign, is shown in Fig. 1. According to the character of variability and values of the mean brightness $\bar{m}$ in Table 1, it may be splitted into several subintervals:
\begin{itemize}
\item
53668-53700:  $\bar{m}$ ranged from 11\fm00 to 11\fm18, and this may be characterized as the ``pre-outburst" stage. However, one may note a slow brightening from 52682 to 53700 with a best fit slope $\dot{m}={\rm d}m/{\rm d}t=-0.00624(19)$ mag/day.
Hereafter the numbers in brackets show the error estimate in units of the last digit.
Another important characteristic of outbursts is a ``decay time" $T_{decay}={\rm d}t/d{\rm m}$ for the descending branch of the outburst (cf. Bailey 1975, Warner 1995), and, similarly, a ``rise time". $T_{rise}=-{\rm d}t/d{\rm m}$ for the ascending branch. According to the properties of the least-square linear fits, $|\dot{m}T_{rise}|=\rho^2,$ where $\rho$ is the correlation coefficient between time $t$ and brightness $m.$
For this time interval of our observations, $T_{rise}=37.0\pm1.1$ days/mag.
\item
53700-53714: "rise to the outburst", the star brightened  from $\bar{m}=11\fm09$ to 10\fm26 during 14 days, $T_{rise}=16.0\pm0.1$ days/mag, $\dot{m}=-0.0584(4).$ The slope at the ascending branch is by 9.4 times larger than in the ``pre-outburst" stage. The shape of the rise is noticeably linear, with a large correlation coefficient $\rho=-0.9668(65).$ The slope is smaller than that for ourburst of dwarf novae by an order of magnitude (cf. Warner 1995), thus indicating a different nature (possibly an increase of the mass transfer rate).
\item
53712-53721: ``top of the outburst", two nights 53712, 53713 from 5 may belong both to the ``rise" and the ``top", as the brightness continued to increase, whereas the value for these two nights fits to the range characteristic for the ``top" . The mean value is $\bar{m}=10\fm341(3).$
\item
53721-53723: ``upper limit for decay", During two days between these nights, the brightness decreased by 0\fm34. Because of bad weather in the intermediate night, the time interval of 2 days is an upper limit for the total duration of decay.
\item
53727-53800: ``post outburst slow brightening", the mean slope is $\dot{m}=-0.00288(5)$ mag/day, the values of $\bar{m}$ for individual  nights may differ from the corresponding linear fit up to $\approx0\fm1.$
\end{itemize}

The observed light curves are shown in Fig. 2. They show a variety of shapes and types of variability. Sometimes the ``$3.2-$ hour variability is prominent (e.g. 53727, 53741-53762), sometimes this component of variability has a remarkably small amplitude (53685, 53698). The amplitude of the quasi-periodic oscillations (QPO), which were also previously reported by other authors, varies from night to night and also within one run.

\begin{figure}
\psfig{file=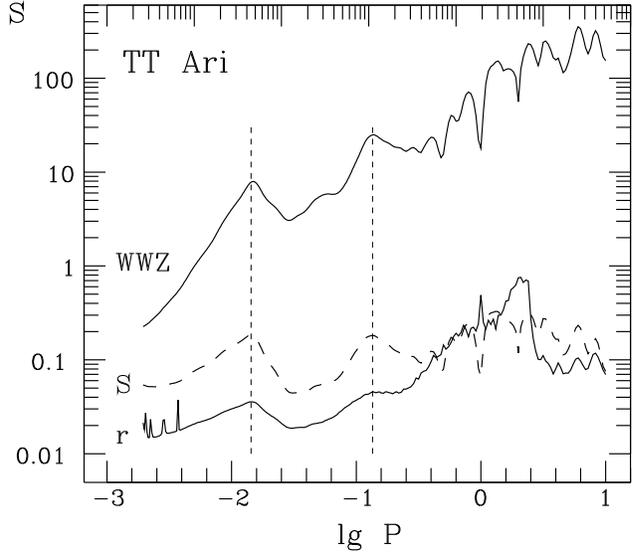} \label{f3} \caption{The ``wavelet
periodograms": test functions used $S(P)$, $r(P),$ WWZ for a
complete set of observations (see Andronov 1998 for a detailed
description). The vertical lines mark the positions of the highest
peaks at $S(f),$ which correspond to $P=0\fd0143=20.6$ minutes and
$P=0\fd135=3\fh24.$}
\end{figure}

\begin{figure*}
\psfig{file=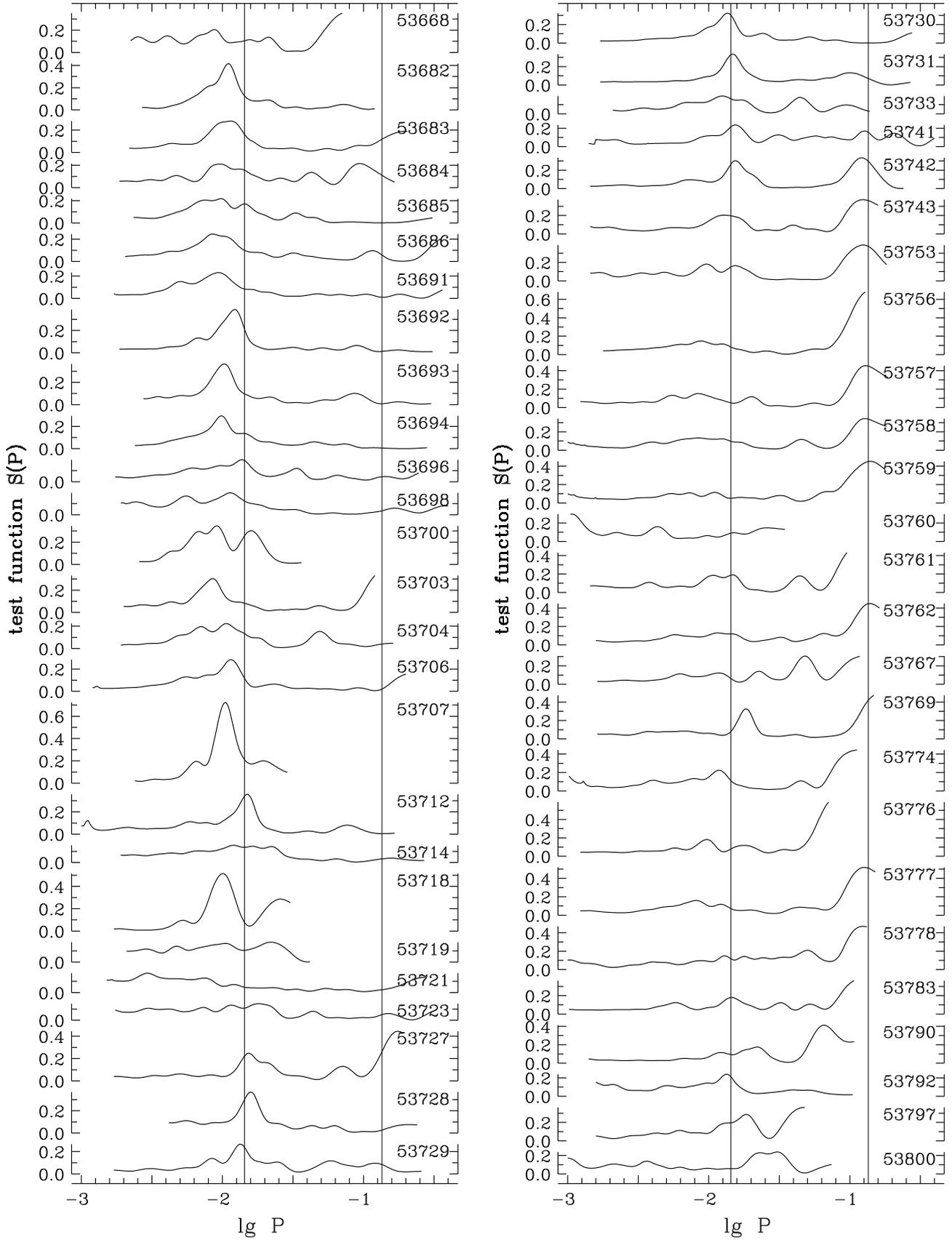} \label{f4} \caption{The individual wavelet
periodograms $S(P)$ for all nights of observations. To avoid bias,
the curves are shown in a range of trial periods from
$3\Delta_{min}$ to $t_e-t_s$ (see Andronov 1998 for a detailed
description).  The vertical lines mark the positions of the highest
peaks of $S(f)$ for the complete set of observations, which
correspond to $P=0\fd0143=20.6$ minutes and $P=0\fd135=3\fh24.$ The
5-digit ``legend Julian date" LJD is shown near each graph. }
\end{figure*}

For numerical parametrization of the observed variability, we used several complementary methods of time series analysis - periodogram, wavelet analysis, running sine and running parabola approximation, O-C analysis, which are described below.

\begin{figure*}
\psfig{file=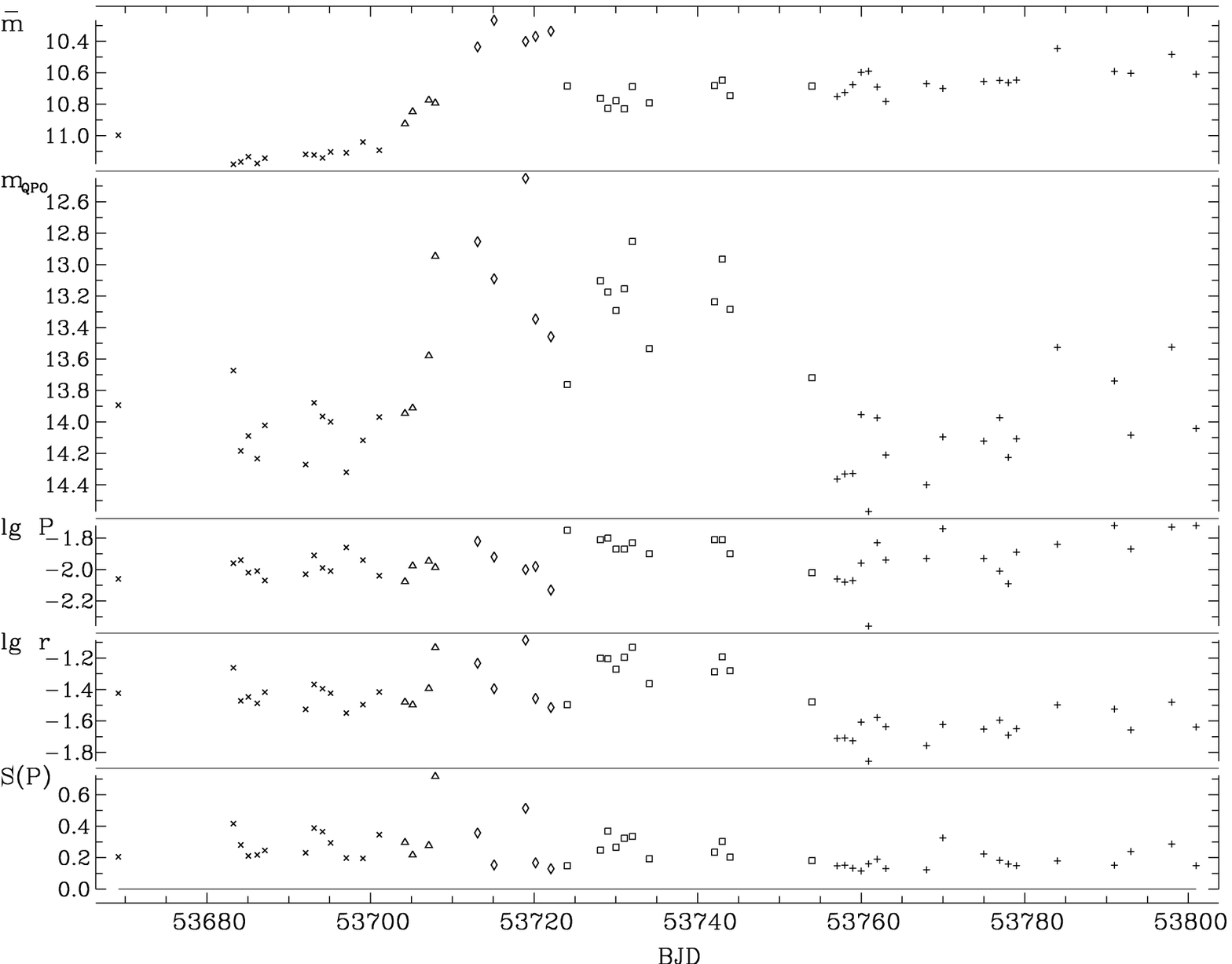} \label{f5}
\caption{Dependence on time of the
characteristics of individual runs: sample mean magnitude $\bar{m},$
effective magnitude of the source of the QPOs $m_{QPO},$ logarithm
of period $\lg P$ of QPOs, logarithm of semi-amplitude $\lg r$ of
QPOs. and the corresponding test function $S(P).$ Each point
represents one night. }
\end{figure*}

\begin{figure*}
\psfig{file=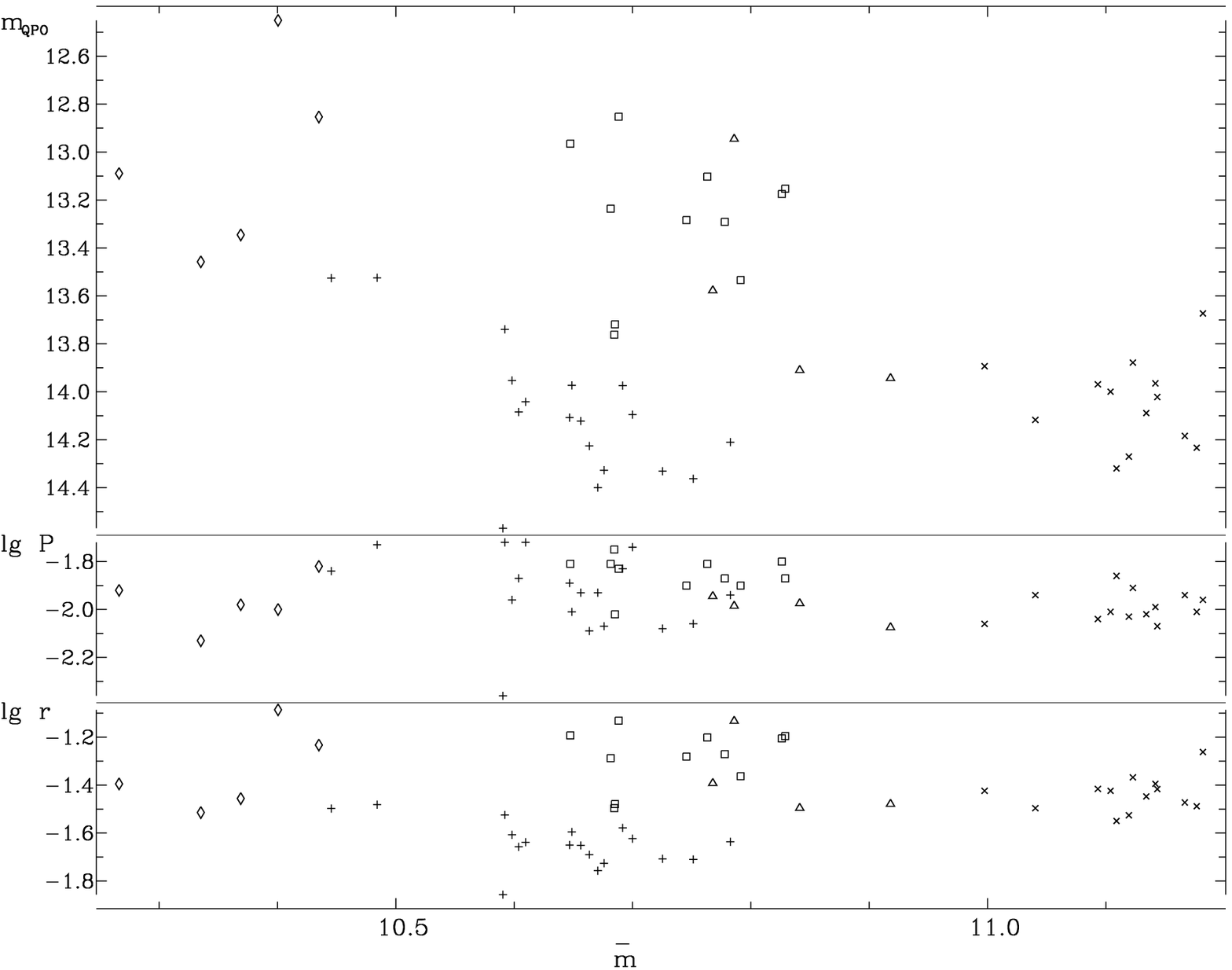} \label{f6} \caption{Dependence on $\bar{m}$ of
the characteristics of individual runs: effective magnitude of the
source of the QPOs $(m_{QPO}),$ logarithm of period $(\lg P)$ of
QPOs, logarithm of semi-amplitude $(\lg r)$ of QPOs.
Each point represents one night.
}
\end{figure*}

\subsection{Wavelet Analysis}

The wavelet analysis is one of the powerful methods to study signals with variable periods and amplitudes (cf. Daubechies 1988).
For the ```wavelet periodograms", we used the test functions $S(P),$  $r(P)$ (Andronov 1998a) and WWZ (Foster 1996), which are specially elaborated for a case of finite signals with possibly non-regular arguments. Their statistical properties were studied by Andronov (1998b). These test functions are complementary, although the peaks at them correspond to effective values of the period. Their physical sense may be briefly explained as follows (see Andronov (1998ab) for a detailed description): $r(P)$ is an effective amplitude (which is equal to a semi-amplitude, if the signal is purely sinusoidal); $S(P)$ is a weighted mean of the squared correlation coefficient, so $S(P)\approx0$ corresponds to a bad periodic fit, and  $S(P)=1$ shows that the period is correct (if the analyzed signal is purely sinusoidal); WWZ is the ``energy" signal-to-noise ratio, so the false alarm ``probability" FAP$\approx \exp(-{\rm WWZ}).$

These test functions are shown in Fig.3 as a function of trial period $P.$ One may note a large scatter of the semi-amplitude $r(P)$ at small values $\lg P$ $(\le -2.4$ and $\ge 0.0).$ This is an effect of discretness of times of observions, which prevents $r(P)$ to be the best test function among the three listed above. The test function $S(P)$
shows highest peaks at $P=0\fd0143=20.6$ minutes $(\lg P=-1.844)$ and $P=0\fd135=3\fh24$ $(\lg P=-0.87).$ The corresponding values of $S(P)$ (the mean contribution of variability at this effective period to the total variance of the signal) are remarkably similar: 0.184 and 0.182. The values of WWZ at these two best periods are equal to 8.0 and 25.0, respectively, so both suggested types of variability are statistically significant. The mean semi-amplitudes of the period are 0\fm0357 and 0\fm0456. The peak at $\lg P\approx0.2$ occurs at the period, which is comparable with distance between the subsequent nights, so the effective number of points is small. So we do not interpret this timescale to be of physical nature of the object.

To study stability of the parameters of the variability, the ``wavelet periodogram" for individual nights are shown in Fig. 4. The range of trial periods $P$ is restricted to $3\Delta_{min}$ to $t_e-t_s,$ where $\Delta_{min}$ is the time resolution (time step between subsequent observations), and $(t_e-t_s)$ is the duration of observations. Both these parameters are listed in Table 1. The peaks close to either $20.6-$min, or $3.2-$ hour show very significant night-to-night changes both in height and the position. 

The period determination of the $3.2-$ hour variability will be discussed in section 3.6, as it exceeds the duration of many nightly runs of our observations and consequently should not be determined from periodograms for so short individual runs because of possible large distortion of the smoothing function.

\subsection{Effective Characteristics of QPOs}

We used the following parameters to characterize the QPOs for individual nights: the period $P,$ semi-amplitude $r(P)$ and ``effective stellar magnitude" $m_{QPO},$ which correspond to the maximum of $S(P)$ in the interval $-2.4\le\lg P\le-1.7.$ This interval was chosen after the examination of the mean and individual ``wavelet periodograms". The parameter $m_{QPO}=\bar{m}+2.5\lg((1+10^{0.4*r})/(1-10^{-0.4*r}))$ is the magnitude estimate of the source of QPOs. The physical sense is the magnitude of the source of emission, which corresponds to brightening from the stellar magnitude $(\bar{m}+r)$ (invisible) to $(\bar{m}-r)$ (visible).

These characteristics are listed in Table 4 and shown in Fig.5.
The mean semi-amplitudes of variability range from 0\fm0139 (run 53760) to 0\fm0821 (run 53718), i.e. vary by a factor of 5.9. The $\lg P$ ranges from -2.13 to -1.72, except one night 53760 with an outlying value of -2.36 (the run was short, so this value represents only a short time interval and thus is not very accurate).

The time variation of $\bar{m}$ was discussed earlier, with splitting the interval of our observations according to this parameter. Surprisingly, the outburst seen in $\bar{m}$ has a very different shape and duration in $m_{QPO}.$ At the beginning of the outburst (53703, 53704), when the mean brightness $\bar{m}$ started to increase, the semi-amplitude $r$ and $m_{QPO}$ are within their ranges characteristic for the ``pre-outburst" stage. Then the values of $m_{QPO}$ range from 12\fm85 to 13\fm76 (except one short night 53718) for the nights 53706-53753. This stage may be referred to as ``QPO-excited", starts from the outburst rise and lasts at least $30^{\rm d}$ after the end of $\approx10^{\rm d}$ ourburst.

After that, $m_{QPO}$ steeply decreases (i.e. the QPO flux increases) with a best fit slope ${\rm d}m_{QPO}/{\rm d}t=-0.0125(37)$ mag/day which is much larger than that of -${\rm d}\bar{m}/{\rm d}t=0.0037(11)$ mag/day. However, night-to-night variability is significant, the slopes only slightly exceed the $``3\sigma-$ level". The best fit for 18 nights (except the short run 53760) is $m_{QPO}=14\fm05(3)+2.61(42)\cdot(\bar{m}-10\fm64),$ the slope is determined at  much more significant $6.3\sigma-$ level". This argues for a statistically significant relationship between $m_{QPO}$ and $\bar{m}.$ The slope is larger than unity, so with increasing mean flux, also increase: the semi-amplitude (related to arbitrary amplitude of the flux variations), the brightness of the source of QPO (related to the flux of QPO) and the period. The statistical significance of latter relationship at the $3.4\sigma-$ level is seen from the fit $\lg r=-1.628(14)+0.40(11)\cdot(\lg P+1.9).$

The relations present above were derived for the part of the light curve after the outburst of the QPO. The complete diagrams showing pairs of parameters (not shown), exhibit more complex behaviour and several groups with smaller number of points inside.
At Fig.6, the diagrams of $m_{QPO},$ $\lg P$ and $\lg r$ vs $\bar{m}$ are shown. One may note two distinctly different groups at the intermediate brightness interval, which are shown by squares (4-th interval) and crosses (5-th interval). These time intervals are consequent, and are characterized by similar range of $\bar{m}.$ However, they are very different by $\lg r,$ and, consequently, $m_{QPO}.$ Contrary to the 5-th interval, there are no significant statistical dependencies on $\bar{m}$ of any of the parameters $m_{QPO},$ $\lg P$ and $\lg r$.

These relations are interesting for future modeling of accretion disks with slowly increasing brightness.

\subsection{QPO Period Variability During the Run}
Drastic variability of the nightly values of the QPO characteristics is observed. To study this phenomenon, we have used the wavelet test function $S(f)$ (Andronov, 1998ab), as the position of peaks of this function coincides to the underlying period. The dependence of the $\lg P$ on time for all individual nights is shown in Fig. 7. According to stability of the periods during the run, one may distinguish between nights with small period variations (e.g. 53712) and multiple switches (e.g. 53691). Such apparent switches at the $P(t)$ diagrams may be explained by aperiodic flares or dips, which occasionally change the interval between the brightness maxima or minima, respectively. One may neglect switches at the borders of observations, as the trial fit used in the wavelet analysis is filled with observations very asymmetrically, what causes larger statistical errors and apparent rise or decay of smoothing signal. Examples of such jumps of frequency for the ``negative superhump" state in 1989 were reported by Kraicheva et al. (1999).

\subsection{Search for Possible Periodic Components}

The presence of QPOs does not mean that there are no truly periodic variations, e.g. related to the spin period of the white dwarf. Similar situation is present in the magnetic dwarf nova DO Dra, where the new phenomenon - ``transient periodic oscillations" hides the spin variability (Andronov et al., 2008). For this purpose, we have used the program "Fo", which realizes the periodogram analysis based on the least squares cosine fit (Andronov 1994). The test function $S(f)$ is a square of the correlation coefficient between the ``observed" and ``computed" (using the least squares sine fit for a trial frequency) values. Statistical significance of the peak increases with its height (see Andronov (1994) for description of statistical properties of this test function).

To avoid a bias of the periodogram caused by significant low frequency ``3\fh2" hour component, it is needed to remove it before the periodogram analysis. Similar to BG CMi (Kim et al., 2005) and DO Dra (Andronov et al., 2008), we have used the ``running sine" (RS) fit, with an initial period $P=0\fd01413$ taken from the wavelet periodogram. Generally, the analysis of the characteristics of this fit (not shown) confirms our previous findings - variability of the period and semi-amplitude. Thus we have used only estimate of the low-frequency trend to compute the residuals of the signal from the smoothing values (O-C).

The test function $S(f)$ for the residuals (O-C) is plotted versus frequency in cycle/day. For reference, the frequency, which corresponds to the best period corresponding to the mean wavelet periodogram, is 69.8 cycles/day. The individual periodograms are shown in Fig. 8. At some nights, the peaks are high and narrow (e.g. 53692, 53731, 53769), which correspond to oscillations, which are stable during the night. Wide peaks are often high (53700, 53707, 53718) and wide, what is consistent with the inverse proportionality of the peak width to the duration of observations. The periodograms show multiple peaks (e.g. 53668), what corresponds to absence of stability of the periods during the nights.  The frequency corresponding to the highest peak, is a good parameter for QPOs. However, due to period changes, the highest peak may not correspond to that at the ``wavelet periodogram" $S(P),$ and the estimate of the semi-amplitude $r$ is much smaller than that for the wavelet (they coincide only, if the signal is strictly periodic). Thus we continue to use estimates from the wavelet analysis. The periodogram for all nights (not shown because of a huge length of the file) shows the highest peak at the frequency $f=68.2712(3)$ (period $P=0\fd0146475(6)=21.0924$ minutes, semi-amplitude $r=0\fm107(6),$ initial epoch
for the maximum brightness $T_{max}=53731.73880(14).$ Although $S(f)=0.02$ is small, i.e. only two per cent of variability may be explained by this periodic component, the ``false alarm probability" FAP=$10^{-56.3}$ is very small, indicating that this variability is statistically significant.

The estimate of the semi-amplitude is much smaller than that from the wavelet analysis, so this peak may be resulted from several nights with QPOs with similar frequency.

\subsection{Superhump Period}

Determination of the period is complicated, as the system exhibits several types of variability. The periodogram analysis needs a preliminary removal of the slow trend, which is not well defined because of the outburst and mean brightness fluctuations. Moreover, the light curve sometimes shows two maxima per preliminary suggested ``3\fh2-hour" period.
Thus we have used the times of prominent minima of the light curve, following an approach used during our previous campaigns (Tremko et al., 1996, Andronov et al., 1999).
For better statistical accuracy, we used the method of ``asymptotic parabola" (Marsakova and Andronov, 1996). The list of 12 minima is presented in Table 5.

To seach for possible periods, we analyzed these 12 times of minima. The program ``PerMin" was used, which realizes the ``method of characteristic events" (Andronov, 1991). In this method, two parameters (period and initial epoch) are being determined, which minimize the r.m.s. deviation of phases from zero. The corresponding dependence of the test function $\sigma(P)$ on the trial period is shown in Fig. 9.

The best fit elements are:
\begin{equation}
T_{min}= {\rm BJD} 2453747.0700(47)+0.13232(5)\cdot E
\end{equation}
The period value is close to that published for previous campaigns: $P=0\fd132959(13)$ (Tremko et al., 1996) and 0\fd133160(4). The difference between the pairs exceeds the statistical errors, but all these values are much smaller, than that in the ``positive superhump" state $P=0\fd14926$ (Skillman et al., 1998) or 0\fd14840 (Andronov et al. 2005). So the star during our set of observations was in the ``negative superhump" state.

The deviations of phases from zero sometimes reach 0.2, what prevents the interpretation of these minima as periodic eclipses. We tried to find a better ephemeris by removing some of the minima, but it has not resulted in better coincidence of phase curves for all nights.

The phase light curves smoothed using the method of ``running parabola" (Andronov, 1997) are shown in Fig. 10.
The time is expressed as a continuously increasing phase computed using the ephemeris (1), i.e. from --1 to 3, not as usually defined from 0 to 1. So practically the brightness variability is shown vs. time (divided by the period) and shifted by some integer number. In TT Ari, the instability of the phase light curve is present not only from night to night, but also from one cycle of variability to another one. So this kind of presentation of the light curve is preferred, like that used for the intermediate polars (e.g. Kim et al., 2005).

 The filter half width is $\Delta t=0\fd05,$ the value adopted by Tremko et al. (1996) and Andronov et al. (1999). The positions of many minima, which were not so prominent, and thus not listed in the Table 5, are in a good agreement with the ephemeris above. However, for other nights, at phase zero, there may be observed a secondary minimum or even a maximum. The light curves may be classified as ``one-maximum" or ``two-maxima" per period.

This result may be explained by strong variations of the disk luminosity, possibly due to the inhomogeneity.

\begin{figure*}
\psfig{file=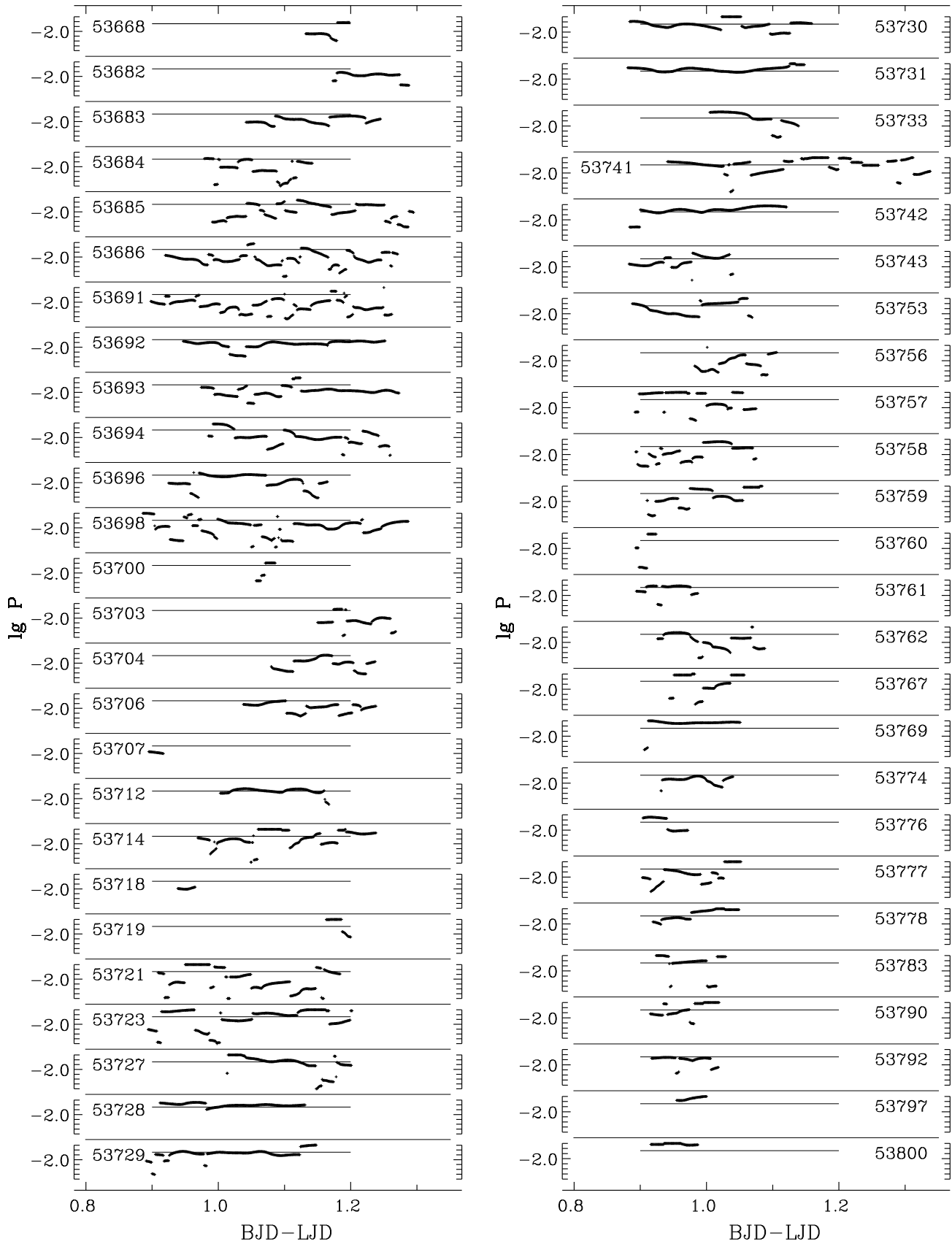}
\label{f7}
\caption{(Available electronically only).Dependence of the local value of the period $P$ of the quasiperiodic oscillations (QPO), which corresponds to a highest maximum of the test function WWZ in the adopted interval $-2.4\le\lg P\le-1.7,$ for the individual nights of observations. The short horizontal lines mark the value of $\lg P=-1.844,$ which corresponds to the maximum of the wavelet periodogram $S(P)$ for all data. The ordinate is marked from -2.4 to -1.7 for all nights.
The 5-digit ``legend Julian date" LJD is shown near each graph. Time is expressed as BJD-LJD, like in Fig. 2.
}
\end{figure*}

\begin{figure*}
\psfig{file=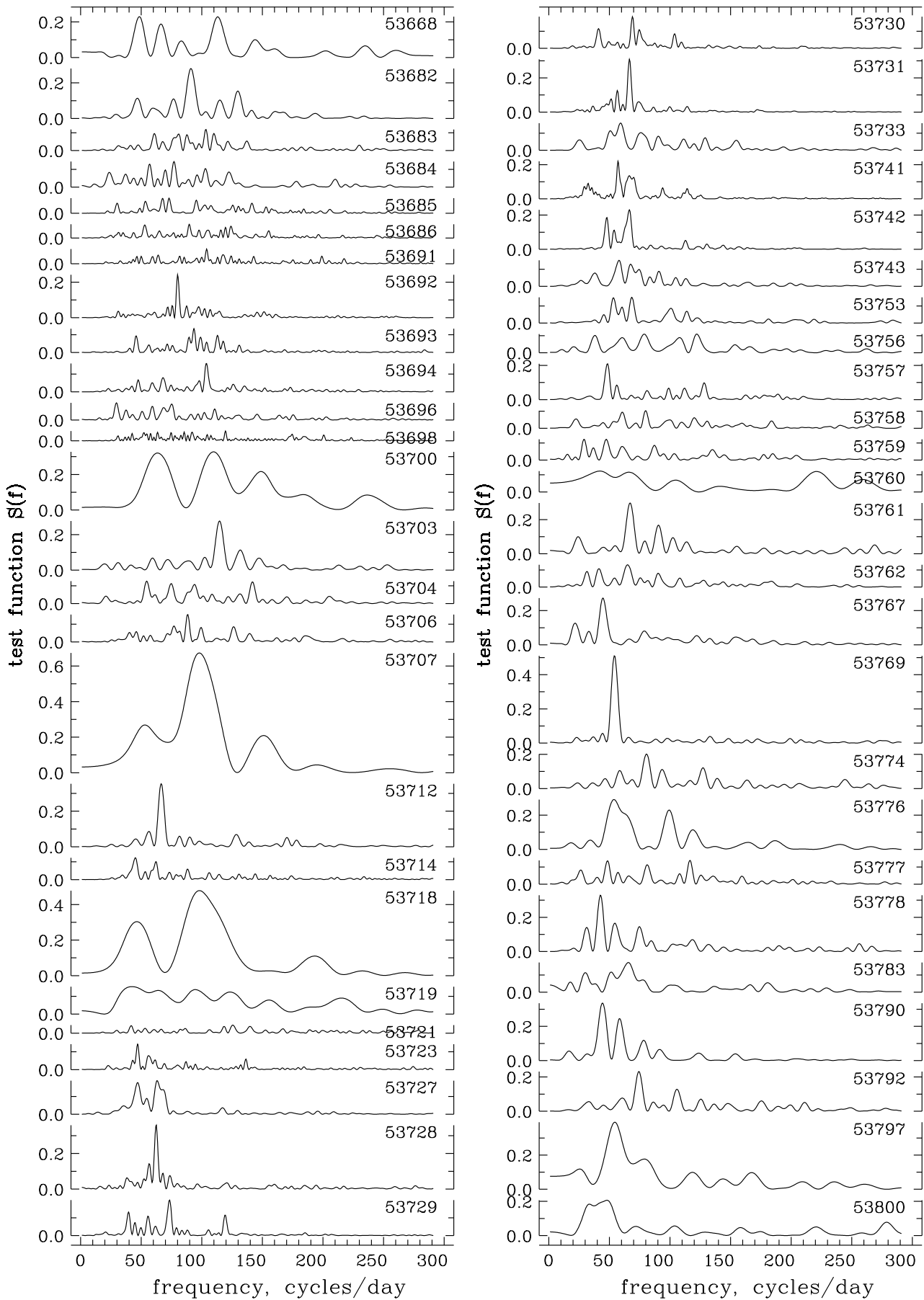}
\label{f8}
\caption{The periodograms $S(f)$ for the individual runs. The 3.2-hour cyclical trend was removed using the ``running sine" method.   The 5-digit ``legend Julian date" LJD is shown near each graph.}
\end{figure*}

\begin{figure}
\psfig{file=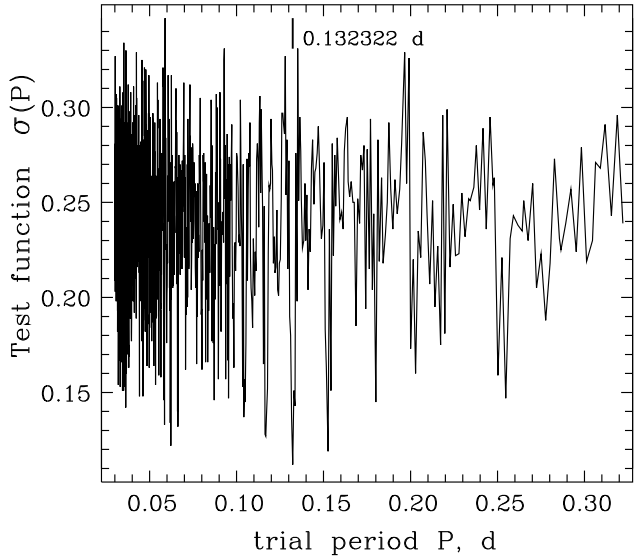} \label{f9} \caption{The dependence of the test
function $\sigma(P)$ for 12 selected moments of minima on the trial
period $P.$ The position of the lowest minimum at P=0\fd132322 is
shown by the short line. }
\end{figure}

\begin{figure*}
\psfig{file=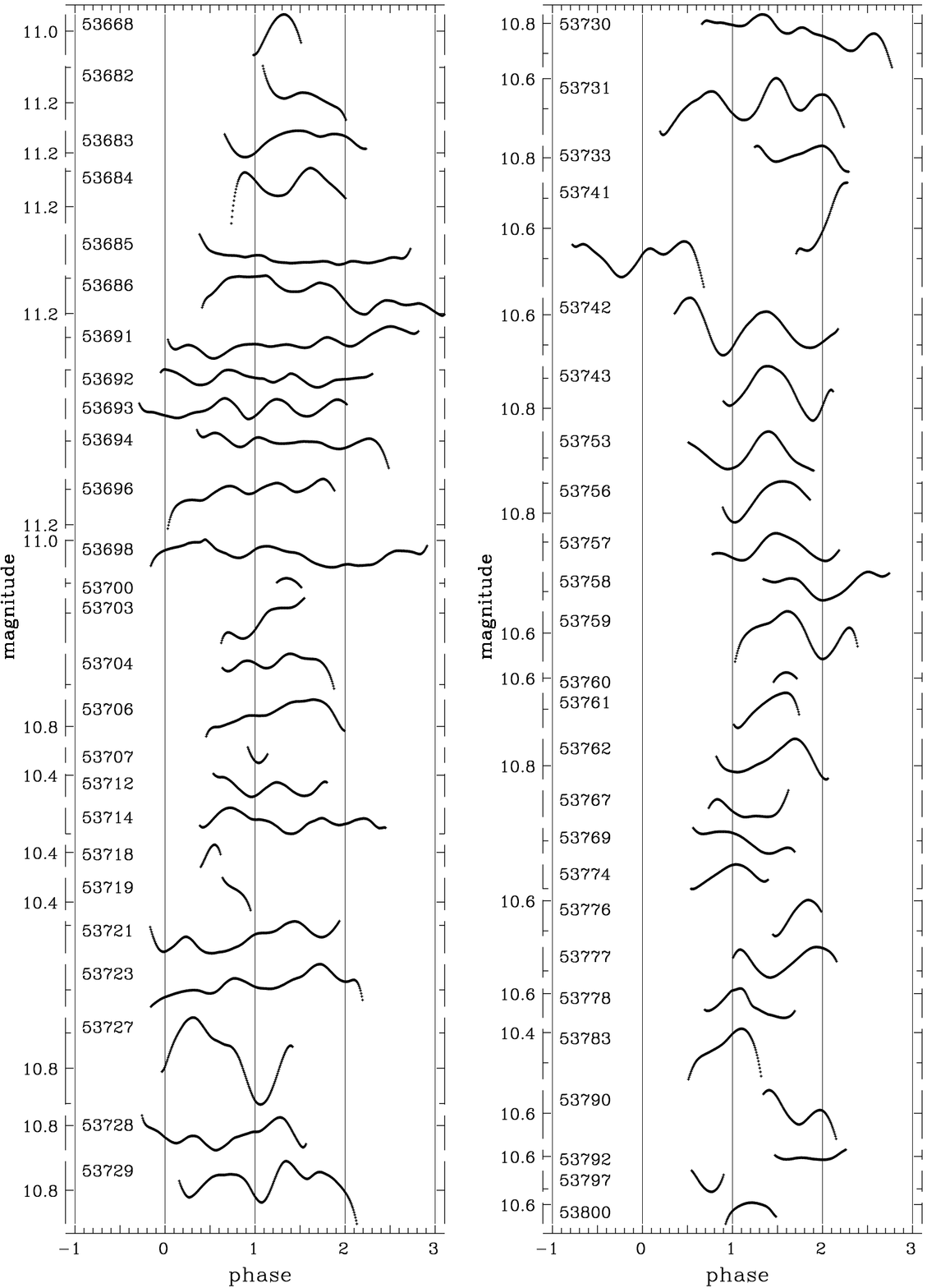}
\label{f10}
\caption{(Available electronically only).Fits for individual nights using the method of running parabola with the filter half-width $\Delta t=0\fd05.$ The time is expressed as a continuously increasing phase computed using the ephemeris (1).
The 5-digit ``legend Julian date" LJD is shown near each graph.
}
\end{figure*}




\section{Conclusions}
Based on the analysis of 51 observational runs (JD 2453669--2453800), which were obtained at the 40-cm telescope of the Chungbuk National University (Korea) and cover 226 hours, the following conclusions are made:
\begin{itemize}
\item
During the period of observations, TT Ari was in the ``negative superhump" state. The preceeding ``positive superhump" lasted from 1996 to 2004.
\item
As in the previous ``negative superhump" states, the light curve is characterized by the ``3.2-" hour variability and a superimposed quasi-periodic oscillations. The shape of the ``superhump" curve is highly variable, showing one-maximum (per period) or two-maxima structure.
\item
The ephemeris for 12 best pronounced ``superhump" minima is $T_{min}=BJD 2453747.0700(47)+0.132322(53)E.$
This is in the range detected in previous studies for the ``negative superhumps". The wide range of the phases of minima (from -0.2 to 0.2)  indicates that these minima may not be eclipses of the source of emission, which is fixed in the rotating frame, but the position may be dynamically variable. Another source of the distortion of the ``superhump" light curve (and, consequently, of the phase scatter) is a strong flickering or aperiodic small-amplitude flares.
\item
The quasi-periodic oscillations (QPO) are present with a mean ``period" of 21.6 min and mean semi-amplitude of 36 mmag. The position and height of the peaks at the ``wavelet periodogram" vary from night to night and sometimes during the night.
\item
The dependence on either time, or the mean brightness, of the magnitude of the source of QPOs, period and amplitude shows that the interval of observations was splitted into 5 parts, showing different characteristics, namely: 1) the ``pre-outburst" stage, 2) the ``rise to outburst", 3) ``top of the outbursts", 4) ``post-outburst QPO" state, 5) ``slow brightening".
The new phenomenon was detected, which we called the ``post-outburst QPO" state. This state is characterized by
a large amplitude of the QPOs for more 30 days (``4") after the outburst (part ``3") lasting 10 days. During both states ``3" and ``4", the amplitude of the QPOs was larger, than in the states ``1" and ``5".
So the photometric behaviour of the system is dependent on the current mean magnitude, but also on the previous state.
\item
The analysis of the pairs of characteristics separately for these groups shows statistically significant correlations only for the state ``5". For this state, with increasing mean brightness, also increase the period, amplitude and brightness of the source of QPOs.
\item
The brightness variations during the present stage of ``negative superhumps" range from 10\fm27 to 11\fm18, whereas the mean brightness in the ``positive superhump" state was only slightly smaller: R$\sim11\fm3$ (Andronov et al. 2004). This means that even small variations in mean brightness (few tenth of magnitude) are sufficient to switch between ``positive" and ``negative" superhumps.
``Positive" superhumps are excited at lower luminosity, than the ``negative" ones.
As TT Ari sometimes exhibits low states to $\approx17^{\rm m},$ it is an interesting observational task to study, which type of variability dominates during low luminosity state.
\end{itemize}

\begin{acknowledgements}
This work was supported by the Korea Research Foundation Grant funded by the Korean Government (MOEHRD, Basic Research Promotion Fund) (KRF-2007-C00121-I00511) and was partially supported by the Ministry of Education and Science of Ukraine.
\end{acknowledgements}


\begin{thebibliography}{}
\bibitem[1991]{A91}Andronov I.L., 1991, KFNT 7, 2, 78
\bibitem[1994]{A94}Andronov I.L., 1994, Odessa Astron. Publ., 7, 49 \mbox{(http://il-a.pochta.ru/oap7\_049.pdf)}
\bibitem[1997]{A97}Andronov I.L., 1997, A\&AS, 125, 207
\bibitem[1998]{A98a}Andronov I.L., 1998a, KFNT, 14, 490
\bibitem[1998]{A98b}Andronov I.L., 1998b, in "Self-Similar Systems", Dubna, 57-70, http://il-a.pochta.ru/dubna.pdf
\bibitem[1999]{A99}Andronov I. L., Arai K., Chinarova L. L. et al.,
1999, AJ, 117, 574
\bibitem[2005]{ABC05}Andronov I.L., Burwitz V., Chinarova L.L. et al.,
2005, IBVS, 5664, 1
\bibitem[2008]{AC08}Andronov I.L., Chinarova L.L., Han W., Kim  Y., Yoon J.-N., 2008, A\&A (in press)
\bibitem[1975]{B75}Bailey J., 1975, J. Brit. Astr. Assoc. 86, 30
\bibitem[1990]{B90}Bianchini A.,
1990, AJ, 99, 1941
\bibitem[1988]{D88}Daubechies I., 1988, Ten Lectures on Wavelets, Cambridge Univ. Press
\bibitem[1998]{E98}Efimov Yu.S., Shakhovskoy N.M., Andronov I.L., Kolesnikov S.V., 1998, BCrAO, 94, 215 
\bibitem[1996]{F96}Foster G., 1996, AJ, 112, 1709
\bibitem[1985]{G85}Goetz W., 1985, IBVS, 2823, 1
\bibitem[2000]{H00}Hellier C., 2000,
NAR 44, 131
\bibitem[2001]{H01}Hellier C.: 2001, Cataclysmic Variable Stars. How and why they vary, Springer Berlin
\bibitem[2007]{Hn07} Henden A., 2007. ftp://ftp.aavso.org/public/calib/ttaribvri.dat
\bibitem[1992]{H92}Hollander A., van Paradijs J., 1992, A\&A, 265, 77
\bibitem[2007]{H07}Hoard D.W., 2007,
http://spider.ipac.caltech.edu/staff/hoard/biglist.html
\bibitem[1984]{H84}Hudec R., Huth H., Fuhrmann B., 1984, Obs, 104, 1
\bibitem[1982]{J82}Jameson R.F., Sherrington M.R., King A.R., Frank, J., 1982, Nature, 300, 152
\bibitem[2005]{KA05b}Kim Y.G., Andronov I.L., Park S.S., Jeon Y.B., A\&A, 2005, 441, 663
\bibitem[1986]{K86}Kozhevnikov V. P., 1986, ATsir, 1455, 5
\bibitem[1981]{K81}Krautter J., Vogt N., Klare G. et al.,
1981, A\&A, 98, 27
\bibitem[1999]{K99}Kraicheva Z., Stanishev V., Genkov V., Iliev L., 1999, A\&A, 351, 607
\bibitem[1999]{L99}Leach R., Hessman F.V., King A.R., Stehle R., Mattei J., 1999, MNRAS,
305, 225
\bibitem[1992]{M92}Massey P., Davis L.E., 1992, A User's Guide to Stellar CCD Photometry with IRAF (Tucson: NOAO)
\bibitem[1980]{M80}
Mardirossian F., Mezzetti M., Pucillo M., et al.,
1980, A\&A, 85 ,29
\bibitem[1994]{AM94b}Marsakova V.I., Andronov I.L., 1996, Odessa Astron. Publ., 9, 127 \mbox{(http://oap09.pochta.ru)}
\bibitem[1980]{N80}Neckel Th., Chini R., 1980, A\&AS, 39, 411
\bibitem[1997]{S97}Semeniuk I., Schwarzenberg-Czerny A., Duerbeck H. et al.,
1987, AcA, 37, 197
\bibitem[1985]{S85}Shafter A.W., Szkody P., Liebert J. et al.,
1985, ApJ, 290, 707
\bibitem[1998]{S98}Skillman D.R., Harvey D.A., Patterson J., et al., 1998, Ap.J., 503, 67L
\bibitem[1987]{S87}Stetson P.B., 1987, PASP, 99, 191
\bibitem[1996]{TAC96}Tremko J., Andronov I.L., Chinarova L.L., et al., 1996, A\&A 312, 121
\bibitem[1995]{W95}Warner B., 1995, Cataclysmic Variable Stars, Cambridge Univ. Press
\bibitem[1996]{W96}Welsh W.F., Martell Ph.J., 1996, MNRAS, 282, 739
\bibitem[1986]{W86}Wenzel W., Bojak W., Critescu C. et al., 1986, Preprint Astron. Inst. Czechoslovak Acad. Sci., 38, 44pp.
\bibitem[2006]{Y06}Yoon J.N., Andronov I. L., Cha S.M., Chinarova L.L., Kim Y.G.,
 2006, ATel, 718, 1
\end{thebibliography}
\end{document}